\documentclass[onecolumn,10pt]{article}
\usepackage[top=.75in, bottom=.75in, left=.75in, right=.75in]{geometry}
\usepackage{amsmath,amsfonts,amscd,amssymb}
\usepackage{graphicx}
\usepackage{epstopdf}
\usepackage{overpic}
\usepackage{cancel}
\usepackage{rotating}
\usepackage{url}
\usepackage{caption}
\usepackage{color}
\usepackage{rotating}
\usepackage{multirow}
\usepackage{wrapfig}
\usepackage{mathtools}
\usepackage{subeqnarray}
\usepackage{setspace}
\usepackage{palatino} 
\usepackage[numbers,sort&compress]{natbib}

\usepackage[bottom,flushmargin,hang,multiple]{footmisc}
\usepackage{lipsum}
\newcommand\blfootnote[1]{%
  \begingroup
  \renewcommand\thefootnote{}\footnote{#1}%
  \addtocounter{footnote}{-1}%
  \endgroup
}

\definecolor{header1}{cmyk}{0,0,0,1}

\usepackage[utf8]{inputenc}

\usepackage[normalem]{ulem}
\usepackage{color}

\title{\vspace{-.45in}{\huge\selectfont \textbf{Perspectives on predicting and controlling turbulent flows through deep learning}}\vspace{-.15in}}

\author{\normalsize{Ricardo Vinuesa$^{1,2*}$}\\
\footnotesize{$^1$ FLOW, Engineering Mechanics, KTH Royal Institute of Technology, Stockholm, Sweden}\\
\footnotesize{$^2$ Swedish e-Science Research Centre (SeRC), Stockholm, Sweden \vspace{-.2in}}
}

\date{}

\begin{document}
\maketitle

\blfootnote{$^*$ Corresponding author: Ricardo Vinuesa (rvinuesa@mech.kth.se)}
\vspace{-.2in}
\begin{abstract}

The current revolution in the field of machine learning (ML) is leading to many interesting developments in a wide range of areas, including fluid mechanics. Here we review recent and emerging possibilities in the context of predictions, simulations and control of fluid flows, focusing on wall-bounded turbulence. A number of important areas are benefiting from ML, and it is important to identify the synergies with the existing pillars of scientific discovery, {\it i.e.} theory, experiments and simulations. It is essential to adopt a balanced approach as a community in order to harness all the positive potential of these novel methods.

{\bf Impact statement.} Fluid mechanics, and more concretely turbulence, is an ubiquitous problem in science and engineering. Being able to understand and predict the evolution of turbulent flows can have a critical impact on our possibilities to tackle a wide range of sustainability problems (including the current climate emergency) and industrial applications. In this piece the potential of novel deep-learning methods to study and predict turbulence is discussed. Another important area where deep learning can help is control, {\it i.e.} the active manipulation of the fluid flow to improve the efficiency of processes ({\it e.g.} reduced drag, increased mixing, etc.).

\vspace{11pt}
    
\noindent\emph{Keywords:} machine learning (ML); deep learning (DL); artificial intelligence (AI); computational fluid dynamics (CFD); experiments; turbulent flows; flow control

\end{abstract}

\section{Predictions in turbulence}\label{sec:pred}

Recent years have witnessed a renewed interest in machine-learning (ML) methods applied to the study of fluid mechanics~\cite{Brunton2020arfm}. The availability of massive amounts of data, together with the possibility of employing more powerful graphics-processing-unit (GPU)-based machines and widely available ML libraries~\cite{tensorflow}, are enabling significant progress at a fast pace. Following some applications to chaotic dynamical systems~\cite{Champion2019pnas}, recent studies start to focus on low-dimensional representations of turbulence~\cite{srinivasan2019predictions} and even wall-bounded turbulent flows~\cite{borreli_et_al_2022}. In these studies the focus is on predicting the temporal dynamics of the flow, a task that can be achieved with quite some success with data-driven methods involving long-short-term-memory (LSTM) networks~\cite{Ref_LSTM} (which are deep-learning architectures capable of exploiting the temporal patterns in the data to perform time-series predictions) and also Koopman-based frameworks with nonlinear forcing~\cite{eivazi2020recurrent}. Other promising approaches for such temporal predictions include reservoir computing~\cite{Magri_Reservoir} (which has been shown to effectively capture extreme events in the time series) and transformers~\cite{heechang} (which have the potential to perform accurate instantaneous predictions over longer time horizons than other data-driven methods). Here it is important to note that, although after a certain time horizon the various data-driven approaches start to deviate with respect to the original time series, the temporal dynamics of the system is well represented as illustrated via {\it e.g.} Poincar\'e maps and Lyapunov exponents~\cite{srinivasan2019predictions}.

Besides temporal predictions, spatial predictions in turbulence can significantly benefit from machine learning. The early work by Milano~\&~Koumoutsakos~\cite{milano_koumoutsakos} showed the feasibility of using deep neural networks to make meaningful predictions in the near-wall region of turbulent channels. Interestingly, these authors showed that by restricting to linear activation functions one can recover the well-known POD (proper-orthogonal decomposition) modes~\cite{lumley}. Another relevant application is non-intrusive sensing, {\it i.e.} the predictions of the flow above the wall based on measurements at the wall. In this spirit, Guastoni {\it et al.}~\cite{guastoni2} showed that it is possible to use convolutional neural networks (CNNs)~\cite{LeCunn_Ref} to effectively exploit the spatial correlations in the turbulence data to perform predictions of the velocity fluctuations based on the two wall-shear-stress fields and the wall pressure. More advanced computer-vision architectures, namely generative adversarial networks (GANs)~\cite{Goodfellow_Ref}, have also been used to obtain robust off-wall predictions based on sparse measurements. This paves the way to obtain methodologies which can be deployed in the context of experiments. Other super-resolution approaches have been proposed by Fukami {\it et al.}~\cite{fukami2019super} and by Yousif {\it et al.}~\cite{yousif_et_al}. By exploiting the similarity between the wall-shear stress and the heat flux at the wall, Kim~\&~Lee~\cite{kim_lee} could leverage the potential of CNNs to predict turbulent heat transfer. Another interesting application of CNNs is the prediction of the flow close to the wall based on information farther away from it~\cite{Ref_Ari}, an approach that could potentially be used for wall modeling.

\section{Simulations of fluid flows}\label{sec:mod}

When it comes to performing computations of fluid flows, ML can help to accelerate direct numerical simulations (DNS), it can improve turbulence modeling and it can help to develop more robust reduced-order models (ROMs)~\cite{vinuesa_brunton}. In the first category, the interesting work by Kochkov {\it et al.}~\cite{hoyer} suggests that it may be possible to accurately perform accurate simulations in coarse meshes. Some more recent work indicates that such an approach to accelerate DNS may also be applied to spectral methods~\cite{hoyer_spectral}. Other approaches to perform turbulence simulations at a reduced cost may be possible thanks to physics-informed neural networks (PINNs)~\cite{raissi2019physics,raissi_et_al}, for instance by effectively solving the Reynolds-averaged Navier--Stokes (RANS) equations~\cite{pinns_pof}. In particular, Eivazi {\it et al.}~\cite{pinns_pof} showed that, given adequate boundary conditions (including those of the Reynolds stresses), the incompressible RANS equations can be efficiently solved, with very good results for the Reynolds stresses, via PINNs. This data-driven approach has also been able to predict the flow from sparse measurements, as described by Sitte~\&~Doan~\cite{sitte_doan}. An alternative way to perform fluid-flow simulations at a reduced computational cost is to reduce the size of the computational domain through data-driven inflow conditions~\cite{syem2,fukami_inflow,heechang} and far-field distributions, which can be obtained {\it e.g.} via Gaussian-process regression~\cite{morita_et_al}.

\begin{figure}[t]
\centering 
\includegraphics[width=0.8\textwidth]{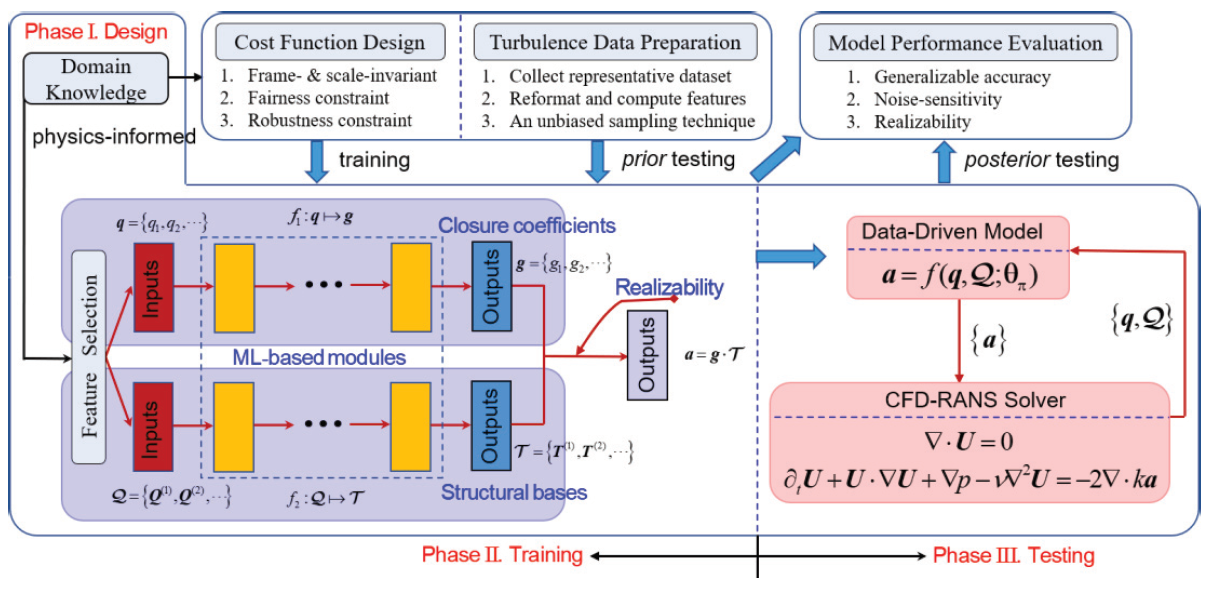}
\vspace{-.1in}
   \caption{Schematic representation of the interpretable machine-learning framework proposed by Jiang~{\it et al.}~\cite{ml_rans} for RANS modeling, which includes the three following phases: (i) design of the framework based on the domain knowledge, (ii) training strategy and (iii) performance assessment. Reprinted from Ref.~\cite{ml_rans}, with permission of the publisher (AIP Publishing).}
   \label{fig:fig1}
\end{figure}

Turbulence modeling is also benefiting from the capabilities enabled by ML. Large-eddy simulations (LES) can rely on accurate subgrid-scale models (SGS) developed by convolutional neural networks (CNNs) together with the eddy-viscosity assumption~\cite{beck_et_al}. In this sense, a promising approach proposed by Novati {\it et al.}~\cite{petros_natmi} relies on using reinforcement learning to determine the coefficient in the Smagorinski model~\cite{smagorinski}. This has led to some success in simulations of turbulent channel. Similar ideas, based on reinforcement learning, have been proposed to develop wall models~\cite{bae_koumoutsakos}; this is an area with interesting potential applications to high-Reynolds-number wall-bounded turbulent flows. Furthermore, RANS simulations have also been improved via ML, more concretely through the development of accurate and robust RANS models. Some of the earlier deep-learning-enabled RANS studies include those by Ling {\it et al.}~\cite{ling2016reynolds} and Wu {\it et al.}~\cite{wu_et_al_2018}, and as anticipated by several perspective papers~\cite{kutz2017deep,duraisamy_et_al}, these data-driven techniques have expanded their application to turbulence modeling. Another data-driven approach to RANS modeling was proposed by Weatheritt~\&~Sandberg~\cite{sandberg2,sandberg1}, who relied on genetic programming to obtain expressions for the Reynolds stresses. The advantage of their work is the fact that they produce interpretable~\cite{rudin,vinuesa_interp} models, {\it i.e.} it is possible to establish an equation relating the inputs and the outputs. In this spirit, Jiang {\it et al.}~\cite{ml_rans} have proposed a deep-learning-based RANS model which is also interpretable, a very important feature of turbulence simulations. Their method, which is shown schematically in Figure~\ref{fig:fig1}, is based on two different neural networks to model the structural and parametric aspects of the turbulent flow.

The last aspect of fluid-flow simulation is the development of ROMs, which can be very helpful in the context of flow control, estimation and optimization. Besides the classical approaches to obtain ROMs, namely proper-orthogonal decomposition (POD)~\cite{lumley,Taira2017aiaa} and dynamic-mode decomposition (DMD)~\cite{Schmid2010jfm} (which both rely on linear algebra), deep learning has been recently used for ROM development by leveraging the non-linearity introduced by the activation functions. Murata {\it et al.}~\cite{murata2020nonlinear} proposed the usage of CNNs, which are capable of exploiting the spatial features in the data, to produce non-linear modal decompositions of fluid flows. Moreover, Fukami {\it et al.}~\cite{fukami_autoencoders} developed an interesting approach for non-linear ROMs based on hierarchical autoencoders. The autoencoder (AE) is a deep-learning model which, through successive application of convolutional filters, can provide a compressed representation of the original data in the latent space. AEs can produce significantly compressed representations of the original data thanks to the employed nonlinearities, but they do not yield modes sorted by energy contribution, and these modes are not orthogonal. These are important properties to provide interpretable and parsimonious ROMs. The hierarchical autoencoder (HAE)~\cite{fukami_autoencoders} relies on the following process:
\begin{enumerate}
    \item An autoencoder with dimension $d=1$ in the latent space is trained, producing one latent vector $\pmb{r}_1$.
    \item Then, another autoencoder with $d=2$ is trained such that $\pmb{r}_1$ is fixed, and a new latent vector $\pmb{r}_2$ is obtained. The contribution of the new vector towards the reconstruction of the original data is smaller than that of the initial one.
    \item Subsequently, a new autoencoder with $d=3$ is trained fixing $\pmb{r}_1$ and $\pmb{r}_2$, producing a new latent vector $\pmb{r}_3$.
    \item This process is repeated until the desired size of the latent space is completed.
\end{enumerate}

\begin{figure}[t]
\centering 
\includegraphics[width=0.9\textwidth]{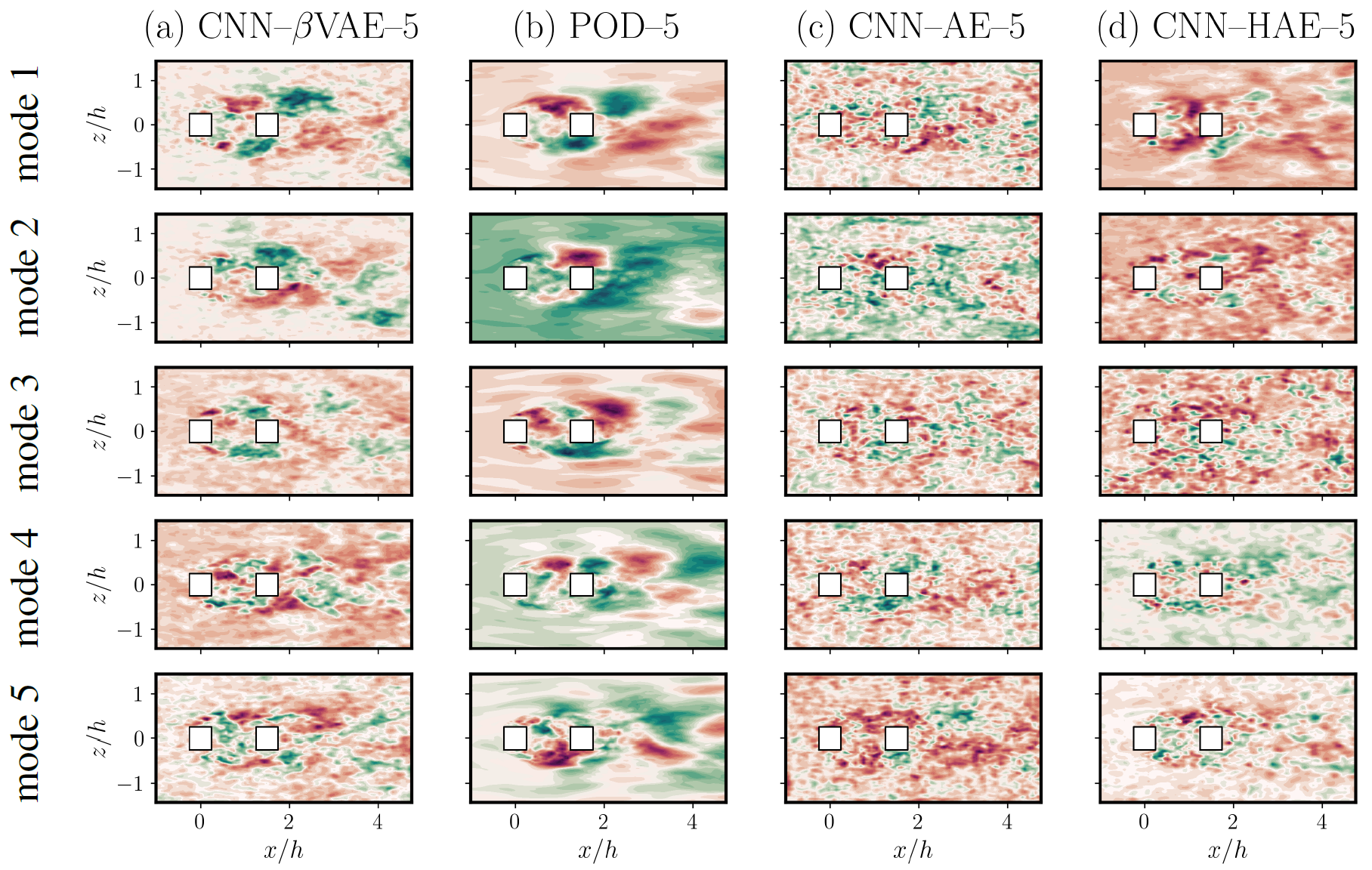}
\vspace{-.1in}
   \caption{The ranked spatial modes obtained from the various approaches for $d=5$. Adapted from Ref.~\cite{ae_modal}, with permission of the publisher (Elsevier).}
   \label{fig:fig2}
\end{figure}

This method produces a series of non-linear AE modes with progressively smaller contribution towards the reconstruction, and Fukami {\it et al.}~\cite{fukami_autoencoders} illustrated its usage in the flow around a two-dimensional cylinder at a Reynolds number $Re=100$ based on cylinder diameter and incoming velocity. The HAE technique was assessed in the turbulent flow between two wall-mounded obstacles~\cite{lazpita_et_al} by Eivazi {\it et al.}~\cite{ae_modal}. While the method works well even in complex turbulent flows, and the modes progressively contribute less towards the reconstruction, their lack of orthogonality affects the interpretability of the ROM. An alternative approach was proposed by Eivazi {\it et al.}~\cite{ae_modal}, and it focused on developing a non-linear orthogonal modal decomposition of the flow while also being able to rank the modes by their contribution towards the reconstruction of the turbulent data. Their method relies on $\beta$-variational autoencoders ($\beta$VAEs), {\it i.e.} a modification of the VAE~\cite{kingma}, an architecture that is receiving attention in the fluid-mechanics community~\cite{maulik2}. In this approach, stochasticity is introduced in the latent space, and a penalization is added to the loss function with the goal of promoting learning the minimum number of nonzero latent variables which are statistically independent. In this way, it is possible to learn a parsimonious and disentangled latent representation of the original data. The $\beta$VAE approach enables recovering almost $90\%$ of the energy from the original data with only 5 modes, whereas the same number of POD modes leads to a recovery of around $30\%$. Furthermore, the $\beta$VAE modes exhibit $99.2\%$ orthogonality, as measured by the cross-correlation matrix, a fact that shows the potential of this method for compact and interpretable ROM development. This point is illustrated in Figure~\ref{fig:fig2}, which shows the first 5 modes from the $\beta$VAE, POD, HAE and a standard AE based on CNNs. The first interesting observation comes from the HAE and standard AE approaches, which exhibit levels of orthogonality below $90\%$, where no distinct patterns or physical phenomena can be really noticed. The first POD modes show the presence of large-scale shedding, which is of course expected in this case, and interestingly this is also identified by the $\beta$VAE modes. In fact, the $\beta$VAE modes also exhibit higher-frequency content in the modes associated with turbulent fluctuations consistent with such shedding, a fact that shows the potential of this approach to significantly compress the data while retaining physical interpretability. Autoencoder-based methods have great potential to develop compact ROMs for turbulent flows, being able to exploit the nonlinear modal reconstruction, as well as potentially interpreting the latent space~\cite{Ref_graham,Ref_magri,Ref_brunton}. Recent work has leveraged transformers~\cite{transformers} to predict the temporal dynamics of the latent space~\cite{solera_et_al,EasyAttention}, effectively producing a ROM to advance the solution in time.   

\section{Control of turbulence}\label{sec:cont}

Data-driven methods for flow control have been extensively used in the literature, for instance exploiting the linear relations in the flow to control transition to turbulence~\cite{Ref_LQ_Fabianne} or forcing at the natural frequency of the shear layer~\cite{Ref_cattafesta} to control separation, a method that can be complented with harmonic resolvent analysis~\cite{Ref_resolvent}. There are however other approaches to flow control based on ML, which may exhibit interesting potential in a number of applications. For instance, Bayesian regression based on Gaussian processes~\cite{Rasmussen} has been used in the context of turbulent-bounday-layer control by Mahfoze {\it et al.}~\cite{Mahfoze}. Another data-driven control approach which has been proved successful in controlling external flows is genetic programming~\cite{li_et_al,minelli_et_al}, which in principle enables exploring various terms in the control law through evolutionary algorithms, and has led to very promising results. One interesting aspect of these approaches is that they allow to assess a larger space of control laws by leveraging the access to large-scale databases, potentially leading to more sophisticated control strategies than the classical ones.

\begin{figure}[t]
\centering 
\includegraphics[width=0.8\textwidth]{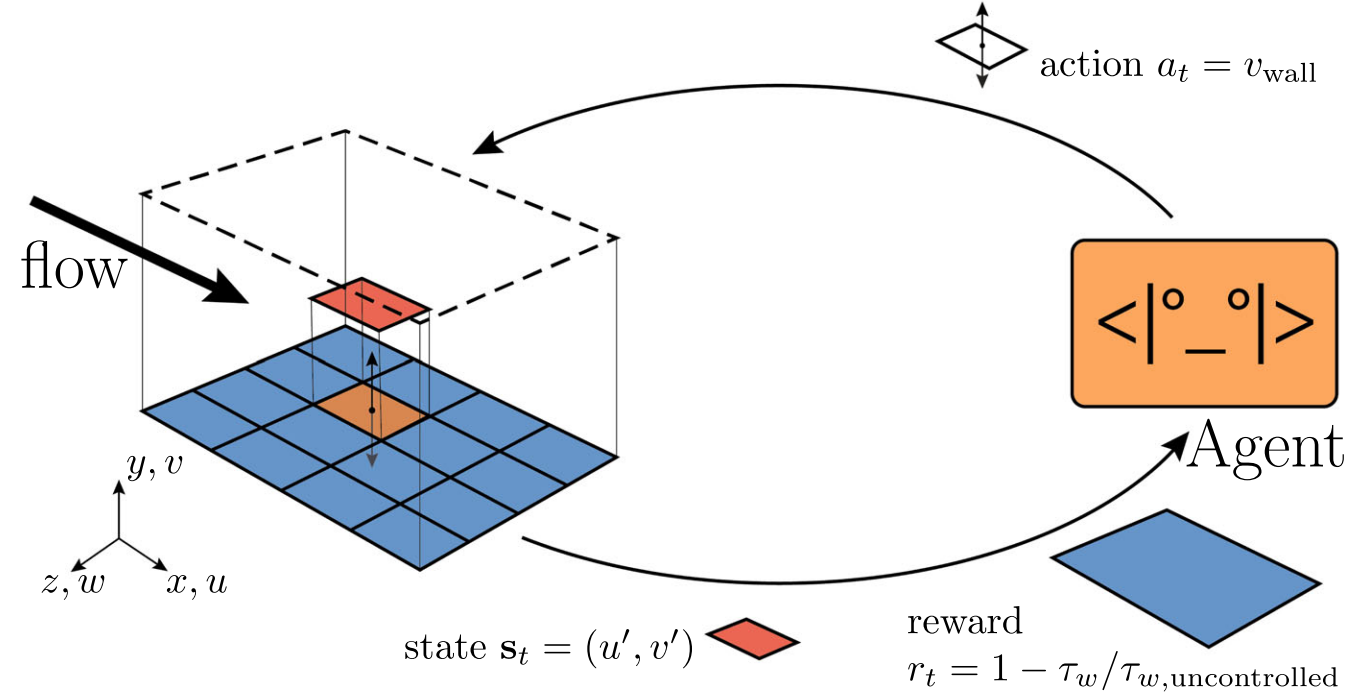}
\vspace{-.1in}
   \caption{Schematic representation of the multi-agent reinforcement learning (MARL) framework proposed by Guastoni {\it et al.}~\cite{guastoni_channel} to control turbulence in channels. The state is defined in terms of the streamwise and wall-normal fluctuations at the sensing plane, the reward is the reduction of wall-shear stress and the actions are blowing/suction at the wall. Reprinted from Ref.~\cite{guastoni_channel}, with permission of the publisher (Springer).}
   \label{fig:marl}
\end{figure}

Another very promising approach to discover novel control strategies is deep reinforcement learning (DRL). In this framework, an agent (the neural network) interacts with an environment (the flow simulator) through actions (the control), thus changing its state. The goal of DRL is to decide the set of actions to take, given the state of the system, in order to maximize a certain reward~\cite{Ref_DRL}. This goal is achieved through iterative interaction with the system by gathering experience on the effect of the actions, a fact that has the potential of producing novel and unexpected control mechanisms in a wide variety of flows. In this sense, one of the first studies to apply DRL for flow control was that of Rabault {\it et al.}~\cite{Ref_Rabault}, who obtained significant drag reduction in the flow around a two-dimensional (2D) cylinder through active control via jets. In this context, Guastoni {\it et al.}~\cite{guastoni_aps} have shown that it is possible to use DRL to reduce the length of a 2D separation bubble. One of their results is the documentation of an improved result from DRL compared with classical periodic control, due to the richer range of frequencies discovered by DRL. Other interesting applications include DRL for control in Couette flow, where the policy learned in a ROM is extended to the full domain~\cite{linot_et_al}, the work on turbulent channels by Sonoda {\it et al.}~\cite{sonoda_et_al} and the work on three-dimensional cylinders by Su\'arez {\it et al.}~\cite{suarez_et_al}. Guastoni {\it et al.}~\cite{guastoni_channel} have recently documented a complete framework based on multi-agent reinforcement learning (MARL), which is illustrated in Figure~\ref{fig:marl}, to perform active flow control in turbulent channels. They document a higher drag reduction through MARL ($30\%$) than that obtained through the classical opposition control ($20\%$)~\cite{choi_et_al}. An extensive account of the potential for flow control from DRL, including turbulent wings, can be found in Refs.~\cite{drl_control,vignon_rev}.
 
\section{Outlook}\label{sec:outlook}

The current revolution in ML, particularly in deep learning, is leading to a number of interesting developments in the context of predicting and controlling fluid flows. It is important to highlight that these methods are not meant to replace any of the current pillars of scientific discovery, {\it i.e.} theory, experiments or simulations; rather, these techniques are enabling to solve concrete problems thanks to an improved predictive power from the available data, and therefore ML can be considered as a fourth complementing pillar instead of a competing approach. In this sense, while very interesting progress is being observed in chaotic systems and simplified fluid flows, the applications to fully-turbulent flows are still scarce. It is in the application of ML methods to large-scale turbulent cases where the most relevant and impactful applications will be found. However, in order to obtain ML models that can effectively be trained on the large datasets from wall-bounded turbulence (on the order of many terabytes of data), it will be necessary to scale up the current algorithms. This will require to address important computer-science questions having to do with the scaling of the algorithms, and their possibility to run concurrently, in situ, with very large simulations. The current trend in high-performance-computing (HPC) facilities, combining both central-processing and graphics-processing units (CPUs and GPUs respectively), further increases the complexity of applying these methods in the future large-scale flow simulations. Despite the challenges, the possibilities in terms of prediction and control are numerous, including in experimental cases~\cite{vinuesa_exp}. Furthermore, already-trained deep-learning models may help to learn novel physical phenomena through interpretability methods~\cite{cranmer_et_al,vinuesa_interp}.

The development of progressively larger databases and models, and the high sensitivity of the results to small hyper parameter choices in deep learning (and more crucially in DRL), make it essential to develop, as a community, better practices in terms of benchmark cases, open-source code development and data sharing. This will enable reproducibility of the results and a quicker (and more robust) advancement of the field of data-driven scientific discovery in fluid mechanics. An important implication of this is the possibility to develop deep-learning models that can generalize to new cases; transfer learning (for example from low to high Reynolds numbers~\cite{guastoni2}) is critical in this sense, as well as the framework of continual learning~\cite{Ref_continual}, which may enable training a deep-learning model which performs equally well in a wide range of tasks. Moreover, novel experimental design aided by ML is another very promising area of research, including data assimilation.

To conclude, there is great potential in ML for fluid mechanics in many areas, and this paradigm can complement the other existing pillars of scientific discovery. In order to properly develop the field and harness all its potential, it is important to maintain the scientific rigor as a community, keeping an informed optimism about the new developments, and avoiding exaggerated claims on the achievements of these techniques.

\section*{Competing Interest}
The author declares no competing interests. 

\section*{Funding}
RV acknowledges the financial support from the Lundeqvist foundation and the ERC Grant No. ``2021-CoG-101043998, DEEPCONTROL''. Views and opinions expressed are however those of the author only and do not necessarily reflect those of the European Union or the European Research Council. Neither the European Union nor the granting authority can be held responsible for them.

\section*{Data Availability Statement}
Data Availability not applicable to this position paper.

\section*{Acknowledgements}
An initial version of this paper is available via the ERCOFTAC Bulletin 134, March 2023, after invitation by Dr. Anh Khoa Doan.

 \begin{spacing}{.88}
 \setlength{\bibsep}{2.pt}
\bibliographystyle{abbrvnat}
\bibliography{aicfd_bib}

\begin{thebibliography}{75}
\providecommand{\natexlab}[1]{#1}
\providecommand{\url}[1]{\texttt{#1}}
\expandafter\ifx\csname urlstyle\endcsname\relax
  \providecommand{\doi}[1]{doi: #1}\else
  \providecommand{\doi}{doi: \begingroup \urlstyle{rm}\Url}\fi

\bibitem[Abadi~{\it et~al}(2016)]{tensorflow}
M.~Abadi~{\it et~al}.
\newblock Tensorflow: A system for large-scale machine learning.
\newblock \emph{Proceedings of the 12thUSENIX Symposium on Operating Systems
  Design and Implementation (OSDI 16) (USENIX Association, 2016)}, 16:\penalty0
  265--283, 2016.

\bibitem[Bae and Koumoutsakos(2022)]{bae_koumoutsakos}
H.~J. Bae and P.~Koumoutsakos.
\newblock {Scientific multi-agent reinforcement learning for wall-models of
  turbulent flows}.
\newblock \emph{Nature Communications}, 13:\penalty0 1443, 2022.

\bibitem[Bakarji et~al.(2023)Bakarji, Champion, Kutz, and Brunton]{Ref_brunton}
J.~Bakarji, K.~Champion, J.~N. Kutz, and S.~L. Brunton.
\newblock Discovering governing equations from partial measurements with deep
  delay autoencoders.
\newblock \emph{Proceedings of the Royal Society A}, 479:\penalty0 20230422,
  2023.

\bibitem[Balasubramanian et~al.(2023)Balasubramanian, Guastoni, G\"uemes,
  Ianiro, Discetti, Schlatter, Azizpour, and Vinuesa]{Ref_Ari}
A.~G. Balasubramanian, L.~Guastoni, A.~G\"uemes, A.~Ianiro, S.~Discetti,
  P.~Schlatter, H.~Azizpour, and R.~Vinuesa.
\newblock Predicting the near-wall region of turbulence through convolutional
  neural networks.
\newblock \emph{International Journal of Heat and Fluid Flow}, 103:\penalty0
  109200, 2023.

\bibitem[Beck et~al.(2019)Beck, Flad, and Munz]{beck_et_al}
A.~D. Beck, D.~G. Flad, and C.-D. Munz.
\newblock {Deep neural networks for data-driven {LES} closure models}.
\newblock \emph{Journal of Computational Physics}, 398:\penalty0 108910, 2019.

\bibitem[Borrelli et~al.(2022)Borrelli, Guastoni, Eivazi, Schlatter, and
  Vinuesa]{borreli_et_al_2022}
G.~Borrelli, L.~Guastoni, H.~Eivazi, P.~Schlatter, and R.~Vinuesa.
\newblock Predicting the temporal dynamics of turbulent channels through deep
  learning.
\newblock \emph{International Journal of Heat and Fluid Flow}, 96:\penalty0
  109010, 2022.

\bibitem[Brunton et~al.(2020)Brunton, Noack, and Koumoutsakos]{Brunton2020arfm}
S.~L. Brunton, B.~R. Noack, and P.~Koumoutsakos.
\newblock Machine learning for fluid mechanics.
\newblock \emph{Annual Review of Fluid Mechanics}, 52:\penalty0 477--508, 2020.

\bibitem[Champion et~al.(2019)Champion, Lusch, Kutz, and
  Brunton]{Champion2019pnas}
K.~Champion, B.~Lusch, J.~N. Kutz, and S.~L. Brunton.
\newblock Data-driven discovery of coordinates and governing equations.
\newblock \emph{Proceedings of the National Academy of Sciences}, 116\penalty0
  (45):\penalty0 22445--22451, 2019.

\bibitem[Choi et~al.(1994)Choi, Moin, and Kim]{choi_et_al}
H.~Choi, P.~Moin, and J.~Kim.
\newblock Active turbulence control for drag reduction in wall-bounded flows.
\newblock \emph{Journal of Fluid Mechanics}, 262:\penalty0 75--110, 1994.

\bibitem[Cranmer et~al.(2020)Cranmer, Sanchez-Gonzalez, Battaglia, Xu, Cranmer,
  Spergel, and Ho]{cranmer_et_al}
M.~Cranmer, A.~Sanchez-Gonzalez, P.~Battaglia, R.~Xu, K.~Cranmer, D.~Spergel,
  and S.~Ho.
\newblock {Discovering symbolic models from deep learning with inductive
  biases}.
\newblock \emph{34th Conference on Neural Information Processing Systems
  (NeurIPS 2020), Vancouver, Canada. Preprint arXiv:2006.11287}, 2020.

\bibitem[Doan et~al.(2021{\natexlab{a}})Doan, Polifke, and
  Magri]{Magri_Reservoir}
N.~A.~K. Doan, W.~Polifke, and L.~Magri.
\newblock Short-and long-term predictions of chaotic flows and extreme events:
  a physics-constrained reservoir computing approach.
\newblock \emph{Proceedings of the Royal Society A}, 477:\penalty0 20210135,
  2021{\natexlab{a}}.

\bibitem[Doan et~al.(2021{\natexlab{b}})Doan, Polifke, and Magri]{Ref_magri}
N.~A.~K. Doan, W.~Polifke, and L.~Magri.
\newblock Auto-encoded reservoir computing for turbulence learning.
\newblock \emph{International Conference on Computational Science. Springer,
  Cham}, pages 344--351, 2021{\natexlab{b}}.

\bibitem[Dresdner et~al.(2022)Dresdner, Kochkov, Norgaard, Zepeda-N{\'u}ñez,
  Jamie, Smith, Brenner, and Hoyer]{hoyer_spectral}
G.~Dresdner, D.~Kochkov, P.~Norgaard, L.~Zepeda-N{\'u}ñez, L.~Jamie, A.~Smith,
  M.~P. Brenner, and S.~Hoyer.
\newblock Learning to correct spectral methods for simulating turbulent flows.
\newblock \emph{Preprint arXiv:2207.00556}, 2022.

\bibitem[Duraisamy et~al.(2019)Duraisamy, Iaccarino, and Xiao]{duraisamy_et_al}
K.~Duraisamy, G.~Iaccarino, and H.~Xiao.
\newblock Turbulence modeling in the age of data.
\newblock \emph{Annual Review of Fluid Mechanics}, 51:\penalty0 357--377, 2019.

\bibitem[Eivazi et~al.(2021)Eivazi, Guastoni, Schlatter, Azizpour, and
  Vinuesa]{eivazi2020recurrent}
H.~Eivazi, L.~Guastoni, P.~Schlatter, H.~Azizpour, and R.~Vinuesa.
\newblock {Recurrent neural networks and Koopman-based frameworks for temporal
  predictions in a low-order model of turbulence}.
\newblock \emph{International Journal of Heat and Fluid Flow}, 90:\penalty0
  108816, 2021.

\bibitem[Eivazi et~al.(2022{\natexlab{a}})Eivazi, Le~Clainche, Hoyas, and
  Vinuesa]{ae_modal}
H.~Eivazi, S.~Le~Clainche, S.~Hoyas, and R.~Vinuesa.
\newblock {Towards extraction of orthogonal and parsimonious non-linear modes
  from turbulent flows}.
\newblock \emph{Expert Systems with Applications}, 202:\penalty0 117038,
  2022{\natexlab{a}}.

\bibitem[Eivazi et~al.(2022{\natexlab{b}})Eivazi, Tahani, Schlatter, and
  Vinuesa]{pinns_pof}
H.~Eivazi, M.~Tahani, P.~Schlatter, and R.~Vinuesa.
\newblock {Physics-informed neural networks for solving Reynolds-averaged
  Navier--Stokes equations}.
\newblock \emph{Physics of Fluids}, 34:\penalty0 075117, 2022{\natexlab{b}}.

\bibitem[Fabbiane et~al.(2014)Fabbiane, Semeraro, Bagheri, and
  Henningson]{Ref_LQ_Fabianne}
N.~Fabbiane, O.~Semeraro, S.~Bagheri, and D.~S. Henningson.
\newblock Adaptive and model-based control theory applied to convectively
  unstable flows.
\newblock \emph{Applied Mechanics Reviews}, 66:\penalty0 060801, 2014.

\bibitem[François-Lavet et~al.(2018)François-Lavet, Henderson, Islam, and
  Bellemare]{Ref_DRL}
V.~François-Lavet, P.~Henderson, R.~Islam, and M.~Bellemare.
\newblock An introduction to deep reinforcement learning.
\newblock \emph{Foundations and Trends in Machine Learning}, 11:\penalty0
  219--354, 2018.

\bibitem[Fukami et~al.(2019{\natexlab{a}})Fukami, Fukagata, and
  Taira]{fukami2019super}
K.~Fukami, K.~Fukagata, and K.~Taira.
\newblock Super-resolution reconstruction of turbulent flows with machine
  learning.
\newblock \emph{Journal of Fluid Mechanics}, 870:\penalty0 106--120,
  2019{\natexlab{a}}.

\bibitem[Fukami et~al.(2019{\natexlab{b}})Fukami, Nabae, Kawai, and
  Fukagata]{fukami_inflow}
K.~Fukami, Y.~Nabae, K.~Kawai, and K.~Fukagata.
\newblock {Synthetic turbulent inflow generator using machine learning}.
\newblock \emph{Physical Review Fluids}, 4:\penalty0 064603,
  2019{\natexlab{b}}.

\bibitem[Fukami et~al.(2020)Fukami, Nakamura, and
  Fukagata]{fukami_autoencoders}
K.~Fukami, T.~Nakamura, and K.~Fukagata.
\newblock Convolutional neural network based hierarchical autoencoder for
  nonlinear mode decomposition of fluid field data.
\newblock \emph{Physics of Fluids}, 32:\penalty0 095110, 2020.

\bibitem[Goodfellow et~al.(2014)Goodfellow, Pouget-Abadie, Mirza, Xu,
  Warde-Farley, Ozair, Courville, and Bengio]{Goodfellow_Ref}
J.~Goodfellow, J.~Pouget-Abadie, M.~Mirza, B.~Xu, D.~Warde-Farley, S.~Ozair,
  A.~Courville, and Y.~Bengio.
\newblock Advances in neural information processing systems.
\newblock \emph{Journal of Fluid Mechanics}, 27, 2014.

\bibitem[Guastoni et~al.(2021{\natexlab{a}})Guastoni, Ghadirzadeh, Rabault,
  Schlatter, Azizpour, and Vinuesa]{guastoni_aps}
L.~Guastoni, A.~Ghadirzadeh, J.~Rabault, P.~Schlatter, H.~Azizpour, and
  R.~Vinuesa.
\newblock Deep reinforcement learning for active drag reduction in wall
  turbulence.
\newblock \emph{APS Division of Fluid Dynamics Meeting Abstracts},
  A19:\penalty0 007, 2021{\natexlab{a}}.

\bibitem[Guastoni et~al.(2021{\natexlab{b}})Guastoni, G\"uemes, Ianiro,
  Discetti, Schlatter, Azizpour, and Vinuesa]{guastoni2}
L.~Guastoni, A.~G\"uemes, A.~Ianiro, S.~Discetti, P.~Schlatter, H.~Azizpour,
  and R.~Vinuesa.
\newblock Convolutional-network models to predict wall-bounded turbulence from
  wall quantities.
\newblock \emph{Journal of Fluid Mechanics}, 928:\penalty0 A27,
  2021{\natexlab{b}}.

\bibitem[Guastoni et~al.(2023)Guastoni, Rabault, Schlatter, Azizpour, and
  Vinuesa]{guastoni_channel}
L.~Guastoni, J.~Rabault, P.~Schlatter, H.~Azizpour, and R.~Vinuesa.
\newblock Deep reinforcement learning for turbulent drag reduction in channel
  flows.
\newblock \emph{European Physical Journal E}, 46:\penalty0 27, 2023.

\bibitem[Hochreiter and Schmidhuber(1997)]{Ref_LSTM}
S.~Hochreiter and J.~Schmidhuber.
\newblock Deep learning.
\newblock \emph{Neural Computation}, 9:\penalty0 1735, 1997.

\bibitem[Jiang et~al.(2021)Jiang, Vinuesa, Chen, Mi, Laima, and Li]{ml_rans}
C.~Jiang, R.~Vinuesa, R.~Chen, J.~Mi, S.~Laima, and H.~Li.
\newblock An interpretable framework of data-driven turbulence modeling using
  deep neural networks.
\newblock \emph{Physics of Fluids}, 33:\penalty0 055133, 2021.

\bibitem[Kim and Lee(2020)]{kim_lee}
J.~Kim and C.~Lee.
\newblock Prediction of turbulent heat transfer using convolutional neural
  networks.
\newblock \emph{Journal of Fluid Mechanics}, 882:\penalty0 A18, 2020.

\bibitem[Kingma and Welling(2013)]{kingma}
D.~P. Kingma and M.~Welling.
\newblock Auto-encoding variational {Bayes}.
\newblock \emph{Preprint arXiv:1312.6114}, 2013.

\bibitem[Kochkov et~al.(2021)Kochkov, Smith, Alieva, Wang, Brenner, and
  Hoyer]{hoyer}
D.~Kochkov, J.~A. Smith, A.~Alieva, Q.~Wang, M.~P. Brenner, and S.~Hoyer.
\newblock {Machine learning-accelerated computational fluid dynamics}.
\newblock \emph{Proceedings of the National Academy of Sciences}, 118:\penalty0
  e2101784118, 2021.

\bibitem[Kutz(2017)]{kutz2017deep}
J.~N. Kutz.
\newblock Deep learning in fluid dynamics.
\newblock \emph{Journal of Fluid Mechanics}, 814:\penalty0 1--4, 2017.

\bibitem[Lazpita et~al.(2022)Lazpita, Mart\'inez-S\'anchez, Corrochano, Hoyas,
  Le~Clainche, and Vinuesa]{lazpita_et_al}
E.~Lazpita, A.~Mart\'inez-S\'anchez, A.~Corrochano, S.~Hoyas, S.~Le~Clainche,
  and R.~Vinuesa.
\newblock On the generation and destruction mechanisms of arch vortices in
  urban fluid flows.
\newblock \emph{Physics of Fluids}, 34:\penalty0 051702, 2022.

\bibitem[LeCun et~al.(2015)LeCun, Bengio, and Hinton]{LeCunn_Ref}
Y.~LeCun, Y.~Bengio, and G.~Hinton.
\newblock Deep learning.
\newblock \emph{Nature}, 521:\penalty0 436--444, 2015.

\bibitem[Ling et~al.(2016)Ling, Kurzawski, and Templeton]{ling2016reynolds}
J.~Ling, A.~Kurzawski, and J.~Templeton.
\newblock Reynolds averaged turbulence modelling using deep neural networks
  with embedded invariance.
\newblock \emph{Journal of Fluid Mechanics}, 807:\penalty0 155--166, 2016.

\bibitem[Linot and Graham(2022)]{Ref_graham}
A.~J. Linot and M.~D. Graham.
\newblock Data-driven reduced-order modeling of spatiotemporal chaos with
  neural ordinary differential equations.
\newblock \emph{Chaos}, 32:\penalty0 073110, 2022.

\bibitem[Linot et~al.(2023)Linot, Zeng, and Graham]{linot_et_al}
A.~J. Linot, K.~Zeng, and M.~D. Graham.
\newblock Turbulence control in plane {Couette} flow using low-dimensional
  neural {ODE-based} models and deep reinforcement learning.
\newblock \emph{International Journal of Heat and Fluid Flow}, 101:\penalty0
  109139, 2023.

\bibitem[Lumley(1967)]{lumley}
J.~L. Lumley.
\newblock The structure of inhomogeneous turbulence.
\newblock \emph{Atmospheric turbulence and wave propagation, A. M. Yaglom and
  V. I. Tatarski (eds). Nauka, Moscow}, pages 166--178, 1967.

\bibitem[Mahfoze et~al.(2019)Mahfoze, Moody, Wynn, Whalley, and
  Laizet]{Mahfoze}
O.~A. Mahfoze, A.~Moody, A.~Wynn, R.~D. Whalley, and S.~Laizet.
\newblock Reducing the skin-friction drag of a turbulent boundary-layer flow
  with low-amplitude wall-normal blowing within a {Bayesian} optimization
  framework.
\newblock \emph{Physical Review Fluids}, 4:\penalty0 094601, 2019.

\bibitem[Maulik et~al.(2020)Maulik, Fukami, Ramachandra, Fukagata, and
  Taira]{maulik2}
R.~Maulik, K.~Fukami, N.~Ramachandra, K.~Fukagata, and K.~Taira.
\newblock Probabilistic neural networks for fluid flow surrogate modeling and
  data recovery.
\newblock \emph{Physical Review Fluids}, 5:\penalty0 104401, 2020.

\bibitem[Milano and Koumoutsakos(2002)]{milano_koumoutsakos}
M.~Milano and P.~Koumoutsakos.
\newblock Neural network modeling for near wall turbulent flow.
\newblock \emph{Journal of Computational Physics}, 182:\penalty0 1--26, 2002.

\bibitem[Minelli et~al.(2020)Minelli, Dong, Noack, and
  Krajnovi\'c]{minelli_et_al}
G.~Minelli, T.~Dong, B.~Noack, and S.~Krajnovi\'c.
\newblock {Upstream actuation for bluff-body wake control driven by a
  genetically inspired optimization}.
\newblock \emph{Journal of Fluid Mechanics}, 893:\penalty0 A1, 2020.

\bibitem[Morita et~al.(2022)Morita, Rezaeiravesh, Tabatabaei, Vinuesa,
  Fukagata, and Schlatter]{morita_et_al}
Y.~Morita, S.~Rezaeiravesh, N.~Tabatabaei, R.~Vinuesa, K.~Fukagata, and
  P.~Schlatter.
\newblock {Applying Bayesian optimization with Gaussian-process regression to
  computational fluid dynamics problems}.
\newblock \emph{Journal of Computational Physics}, 449:\penalty0 110788, 2022.

\bibitem[Murata et~al.(2020)Murata, Fukami, and Fukagata]{murata2020nonlinear}
T.~Murata, K.~Fukami, and K.~Fukagata.
\newblock Nonlinear mode decomposition with convolutional neural networks for
  fluid dynamics.
\newblock \emph{Journal of Fluid Mechanics}, 882:\penalty0 A13, 2020.

\bibitem[Novati et~al.(2021)Novati, de~Laroussilhe, and
  Koumoutsakos]{petros_natmi}
G.~Novati, H.~L. de~Laroussilhe, and P.~Koumoutsakos.
\newblock {Automating turbulence modelling by multi-agent reinforcement
  learning}.
\newblock \emph{Nature Machine Intelligence}, 3:\penalty0 87--96, 2021.

\bibitem[Padovan et~al.(2020)Padovan, Otto, and Rowley]{Ref_resolvent}
A.~Padovan, S.~E. Otto, and C.~W. Rowley.
\newblock Analysis of amplification mechanisms and cross-frequency interactions
  in nonlinear flows via the harmonic resolvent.
\newblock \emph{Journal of Fluid Mechanics}, 900:\penalty0 A14, 2020.

\bibitem[Poletto et~al.(2013)Poletto, Craft, and Revell]{syem2}
R.~Poletto, T.~Craft, and A.~Revell.
\newblock {A new divergence free synthetic eddy method for the reproduction of
  inlet flow conditions for LES}.
\newblock \emph{Flow Turbulence and Combustion}, 91:\penalty0 519--539, 2013.

\bibitem[R. et~al.(2017)R., Noack, Cordier, Bor\'ee, and Harambat]{li_et_al}
L.~R., B.~R. Noack, L.~Cordier, J.~Bor\'ee, and F.~Harambat.
\newblock Drag reduction of a car model by linear genetic programming control.
\newblock \emph{Experiments in Fluids}, 58:\penalty0 103, 2017.

\bibitem[Rabault et~al.(2019)Rabault, Kuchta, Jensen, R{\'e}glade, and
  Cerardi]{Ref_Rabault}
J.~Rabault, M.~Kuchta, A.~Jensen, U.~R{\'e}glade, and N.~Cerardi.
\newblock Artificial neural networks trained through deep reinforcement
  learning discover control strategies for active flow control.
\newblock \emph{Journal of Fluid Mechanics}, 7:\penalty0 62, 2019.

\bibitem[Raghavan and Balaprakash(2021)]{Ref_continual}
K.~Raghavan and P.~Balaprakash.
\newblock Formalizing the generalization-forgetting trade-off in continual
  learning.
\newblock \emph{35th Conference on Neural Information Processing Systems
  (NeurIPS 2021)}, 2021.

\bibitem[Raissi et~al.(2019)Raissi, Perdikaris, and
  Karniadakis]{raissi2019physics}
M.~Raissi, P.~Perdikaris, and G.~E. Karniadakis.
\newblock Physics-informed neural networks: A deep learning framework for
  solving forward and inverse problems involving nonlinear partial differential
  equations.
\newblock \emph{Journal of Computational Physics}, 378:\penalty0 686--707,
  2019.

\bibitem[Raissi et~al.(2020)Raissi, Yazdani, and Karniadakis]{raissi_et_al}
M.~Raissi, A.~Yazdani, and G.~E. Karniadakis.
\newblock {Hidden fluid mechanics: Learning velocity and pressure fields from
  flow visualizations}.
\newblock \emph{Science}, 367:\penalty0 1026--1030, 2020.

\bibitem[Rasmussen(2003)]{Rasmussen}
C.~E. Rasmussen.
\newblock Gaussian processes for machine learning.
\newblock \emph{In Journal fur Urologie und Urogynakologie. Berlin: Springer
  International Publishing. Berlin: Springer International Publishing}, 7,
  2003.

\bibitem[Rudin(2019)]{rudin}
C.~Rudin.
\newblock Stop explaining black box machine learning models for high stakes
  decisions and use interpretable models instead.
\newblock \emph{Nature Machine Intelligence}, 1:\penalty0 206--215, 2019.

\bibitem[Sanchis-Agudo et~al.(2023)Sanchis-Agudo, Wang, Guastoni, Duraisamy,
  and Vinuesa]{EasyAttention}
M.~Sanchis-Agudo, Y.~Wang, L.~Guastoni, K.~Duraisamy, and R.~Vinuesa.
\newblock Easy attention: A simple self-attention mechanism for
  transformer-based time-series reconstruction and prediction.
\newblock \emph{Preprint arXiv:2308.12874}, 2023.

\bibitem[Schmid(2010)]{Schmid2010jfm}
P.~J. Schmid.
\newblock Dynamic mode decomposition of numerical and experimental data.
\newblock \emph{Journal of Fluid Mechanics}, 656:\penalty0 5--28, 2010.

\bibitem[Sitte and Doan(2022)]{sitte_doan}
M.~P. Sitte and N.~A.~K. Doan.
\newblock Velocity reconstruction in puffing pool fires with physics-informed
  neural networks.
\newblock \emph{Physics of Fluids}, 34:\penalty0 087124, 2022.

\bibitem[Smagorinsky(1963)]{smagorinski}
J.~Smagorinsky.
\newblock {General circulation experiments with the primitive equations: I. The
  basic experiment}.
\newblock \emph{Monthly Weather Review}, 91:\penalty0 99--164, 1963.

\bibitem[Solera-Rico et~al.(2023)Solera-Rico, Vila, Gómez, Wang, Almashjary,
  Dawson, and Vinuesa]{solera_et_al}
A.~Solera-Rico, C.~S. Vila, M.~A. Gómez, Y.~Wang, A.~Almashjary, S.~T.~M.
  Dawson, and R.~Vinuesa.
\newblock $\beta$-variational autoencoders and transformers for reduced-order
  modelling of fluid flows.
\newblock \emph{Preprint arXiv:2304.03571}, 2023.

\bibitem[Sonoda et~al.(2023)Sonoda, Liu, Itoh, and Hasegawa]{sonoda_et_al}
T.~Sonoda, Z.~Liu, T.~Itoh, and Y.~Hasegawa.
\newblock Reinforcement learning of control strategies for reducing skin
  friction drag in a fully developed turbulent channel flow.
\newblock \emph{Journal of Fluid Mechanics}, 960:\penalty0 A30, 2023.

\bibitem[Srinivasan et~al.(2019)Srinivasan, Guastoni, Azizpour, Schlatter, and
  Vinuesa]{srinivasan2019predictions}
P.~A. Srinivasan, L.~Guastoni, H.~Azizpour, P.~Schlatter, and R.~Vinuesa.
\newblock Predictions of turbulent shear flows using deep neural networks.
\newblock \emph{Physical Review Fluids}, 4:\penalty0 054603, 2019.

\bibitem[Suárez et~al.(2023)Suárez, Alcántara-Ávila, Miró, Rabault, Font,
  Lehmkuhl, and Vinuesa]{suarez_et_al}
P.~Suárez, F.~Alcántara-Ávila, A.~Miró, J.~Rabault, B.~Font, O.~Lehmkuhl,
  and R.~Vinuesa.
\newblock Active flow control for three-dimensional cylinders through deep
  reinforcement learning.
\newblock \emph{Preprint arXiv:2309.02462}, 2023.

\bibitem[Taira et~al.(2017)Taira, Brunton, Dawson, Rowley, Colonius, McKeon,
  Schmidt, Gordeyev, Theofilis, and Ukeiley]{Taira2017aiaa}
K.~Taira, S.~L. Brunton, S.~Dawson, C.~W. Rowley, T.~Colonius, B.~J. McKeon,
  O.~T. Schmidt, S.~Gordeyev, V.~Theofilis, and L.~S. Ukeiley.
\newblock Modal analysis of fluid flows: An overview.
\newblock \emph{AIAA Journal}, 55\penalty0 (12):\penalty0 4013--4041, 2017.

\bibitem[Vaswani et~al.(2017)Vaswani, Shazeer, Parmar, Uszkoreit, Jones, Gomez,
  Kaiser, and Polosukhin]{transformers}
A.~Vaswani, N.~Shazeer, N.~Parmar, J.~Uszkoreit, L.~Jones, A.~N. Gomez,
  L.~Kaiser, and I.~Polosukhin.
\newblock Attention is all you need.
\newblock \emph{Advances in Neural Information Processing Systems}, 30, 2017.

\bibitem[Vignon et~al.(2023)Vignon, Rabault, and Vinuesa]{vignon_rev}
C.~Vignon, J.~Rabault, and R.~Vinuesa.
\newblock Recent advances in applying deep reinforcement learning for flow
  control: perspectives and future directions.
\newblock \emph{Physics of Fluids}, 35:\penalty0 031301, 2023.

\bibitem[Vinuesa and Brunton(2022)]{vinuesa_brunton}
R.~Vinuesa and S.~L. Brunton.
\newblock Enhancing computational fluid dynamics with machine learning.
\newblock \emph{Nature Computational Science}, 2:\penalty0 358--366, 2022.

\bibitem[Vinuesa and Sirmacek(2021)]{vinuesa_interp}
R.~Vinuesa and B.~Sirmacek.
\newblock {Interpretable deep-learning models to help achieve the Sustainable
  Development Goals}.
\newblock \emph{Nature Machine Intelligence}, 3:\penalty0 926, 2021.

\bibitem[Vinuesa et~al.(2022)Vinuesa, Lehmkuhl, Lozano-Dur\'an, and
  Rabault]{drl_control}
R.~Vinuesa, O.~Lehmkuhl, A.~Lozano-Dur\'an, and J.~Rabault.
\newblock Flow control in wings and discovery of novel approaches via deep
  reinforcement learning.
\newblock \emph{Fluids}, 865:\penalty0 281--302, 2022.

\bibitem[Vinuesa et~al.(2023)Vinuesa, Brunton, and McKeon]{vinuesa_exp}
R.~Vinuesa, S.~L. Brunton, and B.~J. McKeon.
\newblock The transformative potential of machine learning for experiments in
  fluid mechanics.
\newblock \emph{Nature Reviews Physics}, 5:\penalty0 536--545, 2023.

\bibitem[Weatheritt and Sandberg(2016)]{sandberg1}
J.~Weatheritt and R.~D. Sandberg.
\newblock {A novel evolutionary algorithm applied to algebraic modifications of
  the RANS stress-strain relationship}.
\newblock \emph{Journal of Computational Physics}, 325:\penalty0 22--37, 2016.

\bibitem[Weatheritt and Sandberg(2017)]{sandberg2}
J.~Weatheritt and R.~D. Sandberg.
\newblock {The development of algebraic stress models using a novel
  evolutionary algorithm}.
\newblock \emph{International Journal of Heat and Fluid Flow}, 68:\penalty0
  298--318, 2017.

\bibitem[Wu et~al.(2019)Wu, Xiao, Sun, and Wang]{wu_et_al_2018}
J.~Wu, H.~Xiao, R.~Sun, and Q.~Wang.
\newblock {Reynolds-averaged Navier--Stokes equations with explicit data-driven
  Reynolds stress closure can be ill-conditioned}.
\newblock \emph{Journal of Fluid Mechanics}, 869:\penalty0 553--586, 2019.

\bibitem[Wu et~al.(2022)Wu, Meneveau, Mittal, Padovan, Rowley, and
  Cattafesta]{Ref_cattafesta}
W.~Wu, C.~Meneveau, R.~Mittal, A.~Padovan, C.~W. Rowley, and L.~Cattafesta.
\newblock Response of a turbulent separation bubble to zero-net-mass-flux jet
  perturbations.
\newblock \emph{Physical Review Fluids}, 7:\penalty0 084601, 2022.

\bibitem[Yousif et~al.(2022)Yousif, Yu, and Lim]{yousif_et_al}
M.~Z. Yousif, L.~Yu, and H.-C. Lim.
\newblock Super-resolution reconstruction of turbulent flow fields at various
  {R}eynolds numbers based on generative adversarial networks.
\newblock \emph{Physics of Fluids}, 34:\penalty0 015130, 2022.

\bibitem[Yousif et~al.(2023)Yousif, Zhang, Yu, Vinuesa, and Lim]{heechang}
M.~Z. Yousif, M.~Zhang, L.~Yu, R.~Vinuesa, and H.~Lim.
\newblock A transformer-based synthetic-inflow generator for
  spatially-developing turbulent boundary layers.
\newblock \emph{Journal of Fluid Mechanics}, 957:\penalty0 A6, 2023.

\end{thebibliography}
 \end{spacing}
\end{document}